\documentclass[pra,bibnotes,twocolumn]{revtex4}
\usepackage{graphicx}
\begin{document}
\draft
\def\ds{\displaystyle}
\title{ State conversions around exceptional points }
\author{C. Yuce }
\address{Department of Physics, Eskisehir Technical University, Eskisehir, Turkey }
\email{cyuce@eskisehir.edu.tr}
\date{\today}
\begin{abstract}
We study state conversion in parity-time (PT) symmetry broken non-Hermitian two level system. We construct a theory and explain underlying mechanism for state conversion and define adiabatic evolutions in non-Hermitian systems. The adiabatic theorem can be used if initial state is an instantaneous eigenstate in Hermitian systems. We show that the system can adiabatically follow the modulationally stable instantaneous eigenstate sooner or later, regardless of initial state if the system parameters are varied slowly in non-Hermitian systems. We discuss the topological feature of dynamical circling in the parameter space.

\end{abstract}
\maketitle

\section{Introduction}

Exceptional points (EPs), first predicted decades ago by Kato \cite{EP1}, occur in non-Hermitian systems when at least two eigenvalues and the corresponding eigenvectors coalesce \cite{EP1,EP2,EP3}. EPs are not only of theoretical interest, but also of practical significance since they have some interesting effects such as unidirectional transparency \cite{unitran1,unitran2,unitran3}, lasing and anti-lasing in a cavity \cite{unitran4}, enhanced optical sensitivity \cite{sense1,sense2} and stopping of light \cite{listop}, to name a few. EPs have been experimentally realized in various physical systems \cite{EPdeney1,EPdeney2,EPdeney3,EPdeney4,EPdeney5}. In the literature, single order EP in non-Hermitian systems have been mainly investigated and higher order EPs have also attracted attention in recent years \cite{EPmulti1,EPmulti2,EPmulti3,EPmulti4}. EPs are believed to play essential roles in the theory of topological insulators in non-Hermitian systems. The extension of topological phase to non-Hermitian systems is relatively a new subfield of study and has recently attracted great deal of attention \cite{nonh2,nonh3,ndiakl38}. An interesting feature about EPs is the exchanging instantaneous eigenstates with each other in quasi-static limit \cite{EPdeney1}. On the other hand, only one of the eigenstates remarkably dominates at the end of the cycle if the EP is enclosed periodically in parameter space \cite{EPatom,dynexcep2,encircek0,encircek1,encirc1,encirc2a,encirc3a,encirc4a}. The final state at the end of each cycle is shown to be determined only by the encirclement direction, regardless of the initial state. More specifically, reversing the direction of encircling the EP changes the final states. This implies that the standard adiabatic theory does not work in non-Hermitian systems. This interesting chiral behaviour have recently been confirmed in the optical \cite{EPdeney2,encirc1} and optomechanical \cite{optoep} domains. We note that an anti parity-time ($\mathcal{PT}$) symmetric system was shown to have chiral dynamics if the starting and end points are in the parity-time broken phase \cite{encirc1}. Recently, two interesting papers appeared in the literature. In the first one, a robust asymmetric state exchange even away from the EP was shown to happen \cite{sorun1}. This finding challenged the previous belief that enclosing EP is necessary for chiral state conversion. In the second one, the dynamical encircling of an EP was shown both theoretically and experimentally to lead to a nonchiral behavior when the starting point is in the broken phase \cite{sorun2}. This shows that starting point in the parameter space plays an important role in the dynamics. These two papers opens discussion on the role of the EP on chiral state conversion. A theoretical model is needed to understand the underlying physics and reveal the role of EP, starting point and the direction and magnitude of angular speed.\\
In this paper, we construct a theory to explain state conversion in non-Hermitian systems. We discuss adiabatic conditions and explore  state exchange in both low and high frequency regimes. We consider closed trajectories in parameter space and show that state conversions can occur for trajectories either excluding or including exceptional points.  Furthermore, we show that dynamical encircling has topological feature. We define stable and unstable eigenstates in non-Hermitian systems and show that any initial state eventually evolves to stable eigenstates.

\section{State conversion}
 
We consider the following two-level time-dependent non-Hermitian Hamiltonian
\begin{equation}\label{yudj2}
\mathcal{H}=\left(\begin{array}{ccccc}i \gamma(t)+\delta(t) & -1  \\ -1  & -i\gamma(t)-\delta(t)  \end{array}\right)
\end{equation}
where $\ds{\gamma(t)}$ is gain/loss strength, $\delta(t)$ is detuning and the tunneling amplitude is set to $-1$ for simplicity. The state vector can be found using $\ds{\mathcal{H}|\psi(t)>=i\partial_t |\psi(t)>}$, where $\ds{|\psi(t)>=\{a(t),b(t)\}^{T}}$ .\\
Let us start to find the eigenstates analytically for the spacial case where $\ds{\gamma(t)=\gamma_0}$ and $\ds{\delta(t)=\delta_0}$ are constants. In this case, the non-orthogonal unnormalized eigenstates with eigenvalues $\ds{ E_\mp=\mp\sqrt{1+(\delta_0+i\gamma_0)^2}  }$ are given by
\begin{eqnarray}\label{oawipjj2}
\psi_\mp=\left(\begin{array}{cc}  1 \\ \delta_0+i\gamma_0\pm  \sqrt{1+(\delta_0+i\gamma_0)^2}    \end{array}\right)
\end{eqnarray}
We note that $\psi_+$ grows in time while $\psi_-$ decays when the signs of $\delta_0$ and $\gamma_0$ are the same. The opposite is true if they have opposite signs. This can be seen from the imaginary part of the energy eigenvalues for these two states. Of special importance is the exceptional point, which occurs at $\gamma_0={\mp}1$ and $\delta_0=0$. The unique state vector at the exceptional point is given by $\ds{|\psi_{EP}>=\{1,{\mp}i\}^{T}}$ with zero eigenvalue. \\ 
We showed earlier that only the state with higher imaginary part of energy eigenvalue is modulationally (or dynamically) stable \cite{cemmod}. We discussed that any initial state but the dynamically stable eigenstate are unstable against amplitude modulation, which can arise due to unavoidable noises. Let ${\Psi_1}$ and ${\Psi_0}$ be the dynamically stable and unstable eigenstates whose imaginary parts of energy eigenvalues are positive and negative, respectively. Therefore ${\Psi_1}=\psi_+$ and ${\Psi_0}=\psi_-$ when $\mathcal{I}m\{E_+\}>0$ and ${\Psi_1}=\psi_-$ and ${\Psi_0}=\psi_+$ when $\mathcal{I}m\{E_+\}<0$. Below we study time evolution of any arbitrary initial state for the time-independent Hamiltonian, i.e., $\ds{\gamma(t)=\gamma_0}$ and $\ds{\delta(t)=\delta_0}$.

\subsection{Arbitrary initial states}

Assume an arbitrary initial state: $\ds{|\psi(t)>=\{a_0,b_0\}^{T}}$, where $\ds{a_0}$ and $\ds{b_0}$ are complex numbers. Let us firstly discuss our problem qualitatively. Time evolution of an initial state can be studied in four cases: i-) $\ds{\gamma_0={\mp}1}$ and $\ds{\delta_0=0}$, ii-) $\ds{|\gamma_0|<1}$ and $\ds{\delta_0=0}$, ii-) $\ds{\gamma_0}$ and $\ds{\delta_0}$ have the same signs iv-) $\ds{\gamma_0}$ and $\ds{\delta_0}$ have the opposite signs. In the first case, the state at large times is always the exceptional state, $\ds{|\psi(t)|>=\{1,{\mp}i\}^{T}}$, regardless of the initial state. This is because there is only one eigenstate at the EP, and any initial state will eventually be the exceptional state. In the second case, energy eigenvalues are real and any initial state oscillates between $\psi_+$ and $\psi_-$. In the last two cases, the $\mathcal{PT}$ symmetry is broken and the energy eigenvalues are complex whose imaginary part is positive for $\Psi_1$  and negative for $\Psi_0$. Therefore, $\Psi_1$ grows in time while $\Psi_0$ decays. In other words, the state at large times is always $\psi_{+}$ ($\psi_{-}$) if the signs of $\delta_0$ and $\gamma_0$ are the same (opposite to each other). \\
Having discussed the problem qualitatively, let us now perform analytical calculations. If we solve the corresponding equation, we find the 
solution exactly $\ds{|\psi(t)>=\{a(t),b(t)\}^{T}}$
\begin{eqnarray}\label{okjhgs2}
a(t)&=& a_0 \cosh{( \nu t)}  -i\frac{  (\delta_0+i\gamma_0)a_0 -b_0  }{\nu } \sinh{\nu t}\nonumber\\
b(t)&=&b_0 \cosh{( \nu t)}  +i \frac{  a_0+(\delta_0+i \gamma_0) b_0    }{\nu } \sinh{\nu t} 
\end{eqnarray}
where $\ds{\nu=\sqrt{-1-(\delta_0+i\gamma_0)^2}}$ with $\ds{\nu^2=-E_{\mp}^2}$. We are interested in large time behaviour.  A rigorous way to identify the state at large times is to calculate the ratio $\ds{\frac{b(t)}{a(t)}  }$. As a special case with $\ds{\gamma_0=\mp1     }$ and $\delta=0$, we obtain $\ds{ a(t)\sim  i( b_0\pm  i a_0   )~t }$ and $\ds{ b(t)= i (  a_0{\mp}i   b_0  )  ~ t     }$ for large values of time. Then the ratio at large times becomes $\ds{\frac{b(t)}{a(t)}\sim {\mp} i  }$. This implies that an arbitrary initial state will eventually be the exceptional state at $\ds{\gamma_0=\mp1     }$. Consider next $\ds{|\gamma_0|<1}$ and $\ds{\delta_0=0}$. In this case, $\ds{\nu}$ becomes purely imaginary and $\ds{a(t)}$ and $b(t)$ change periodically in time. The system oscillates between the two eigenstates and the oscillation depends on the initial values $a_0$ and $b_0$. For the last two cases, we can derive the ratio for large values of time. If $\ds{\gamma_0}$ and $\ds{\delta_0}$ have the same signs, then the ratio at large times becomes
\begin{equation}\label{rfghnm2}
\frac{b(t)}{a(t)} \sim \delta_0+i\gamma_0-\sqrt{1+(\delta_0+i\gamma_0)^2}  
\end{equation}
which is independent of $a_0$ and $b_0$. This means that the state always becomes $\psi_+$, regardless of the initial state as can be easily seen by inspecting Equ. (\ref{oawipjj2}). Conversely, if $\ds{\gamma_0}$ and $\ds{\delta_0}$ have the opposite signs, then the ratio at large times becomes
\begin{equation}\label{rfghnm3}
\frac{b(t)}{a(t)} \sim\delta_0+i\gamma_0+\sqrt{1+(\delta_0+i\gamma_0)^2}  
\end{equation}
In this case, the state always evolves to $\psi_-$ no matter what the initial state is. These two ratios show us that the system's behavior at large times does not depend on the initial values $a_0$ and $b_0$ in the $\mathcal{PT}$-broken region. To this end, we note that the formulas (\ref{rfghnm2}) and (\ref{rfghnm3}) are replaced under time reversal $\ds{t{\rightarrow}-t}$ in Equ. (\ref{okjhgs2}). \\
Our analytical treatment shows that any initial state collapses into the dynamically stable eigenstate $\Psi_1$  at large times in the $\mathcal{PT}$-broken region. A question arises. How long does it take for the system to be in the eigenstate $\Psi_1$ for a given initial state? We can now define {\it{relaxation time $\ds{t_c}$}} as the time required for being in the dynamically stable eigenstate. It is a time scale that characterizes the state conversion of an initial state into the least decaying eigenstate. We say that the relaxation time depends on the initial conditions and decreases with increasing imaginary part of energy eigenvalues.
\begin{figure}[t]
\includegraphics[width=4.25cm]{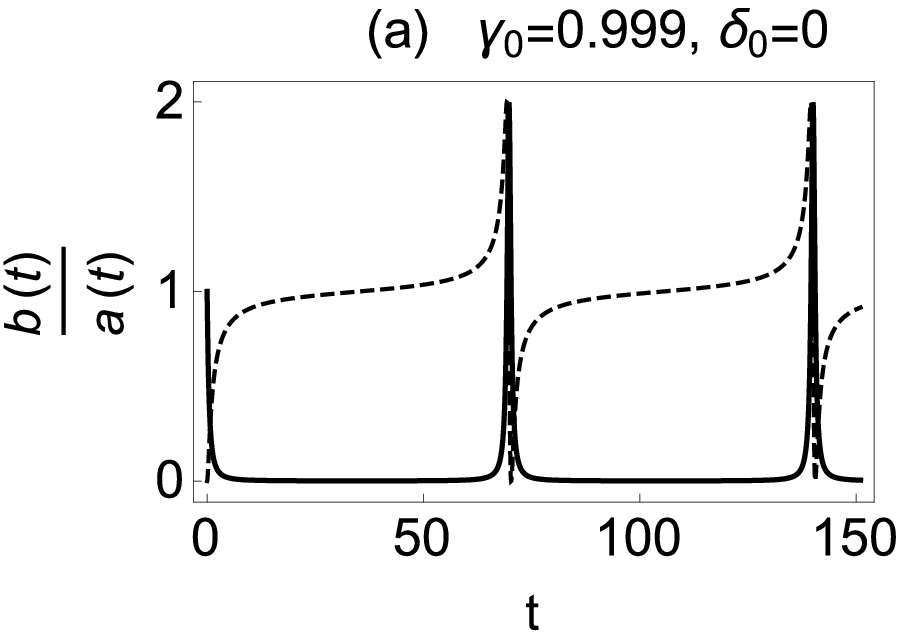}
\includegraphics[width=4.25cm]{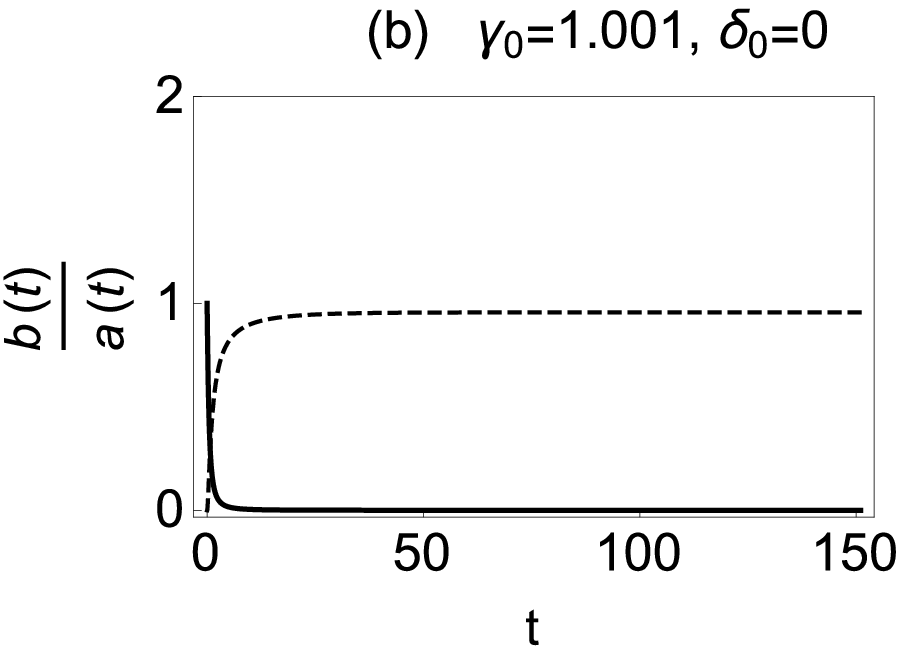}
\includegraphics[width=4.25cm]{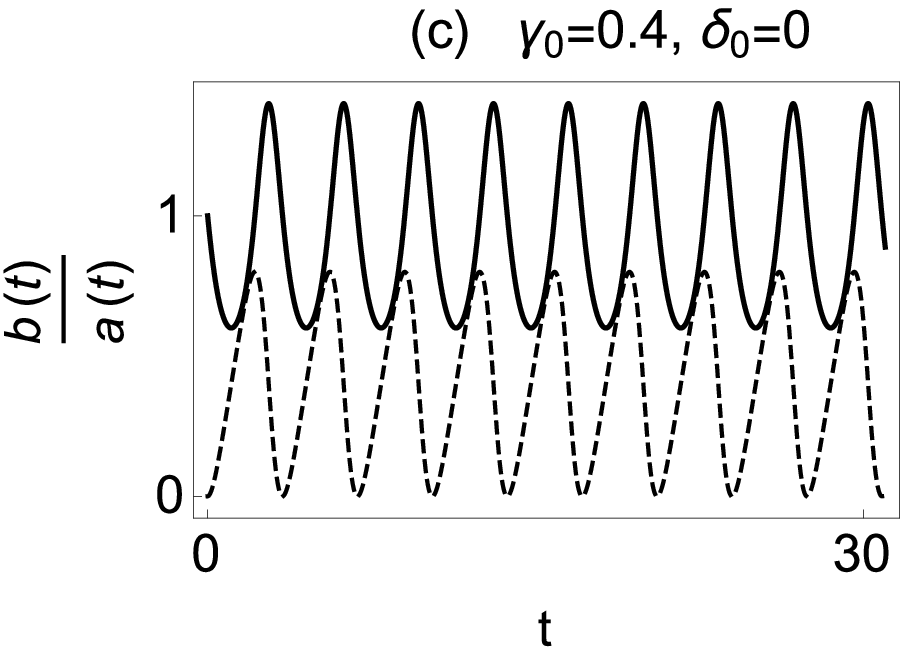}
\includegraphics[width=4.25cm]{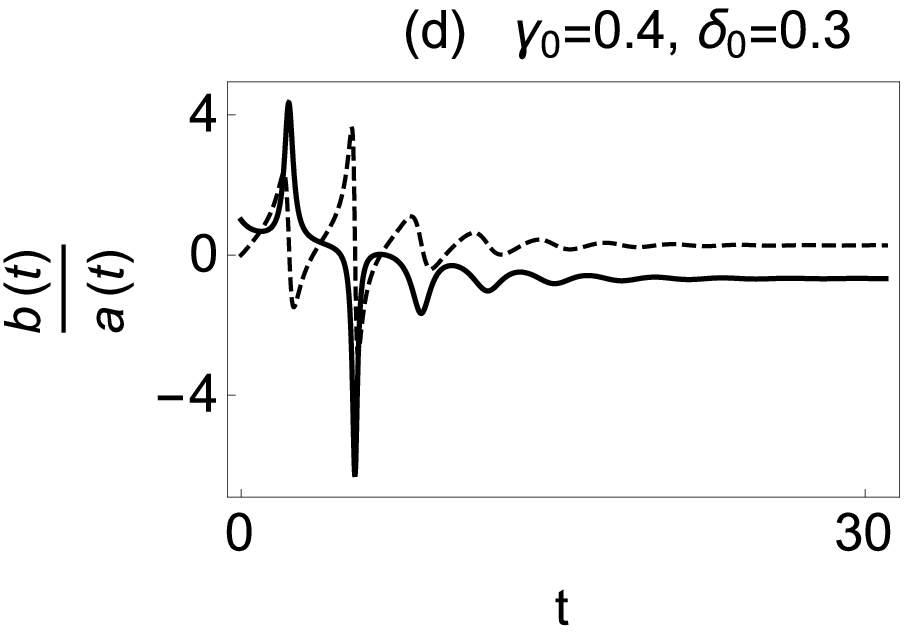}
\caption{ The real (thick) and imaginary (dashed) parts of the ratio $\ds{b(t)/a(t)}$ for the initial state $\ds{ |\psi(t=0)>=\{1,1\}^{T}  }$ for the time-independent Hamiltonian. The system makes oscillations between $\psi_+$ and $\psi_-$ with the period $\ds{T=2\pi/E_+}$ when $|\gamma_0|<1$ and $\delta=0$ (a, c). If $\gamma_0$ is very close to $1$ as in (a), then $\psi_+$ becomes dominant in the oscillation. The state at large times is always $\psi_+$ when $\gamma_0$ and $\delta_0$ have the same signs. }
\end{figure}\\
The Fig.1 plots the real and imaginary parts of the ratio $b/a$ for various values of $\gamma_0$ and $\delta_0$. As can be seen from (a) and (c), the system makes oscillations with the period $\ds{T=2\pi/E_+}$ when $|\gamma_0|<1$ and $\delta=0$. If $\gamma_0$ is very close to $1$ as in (a), then the oscillation is interesting. The system wants to be in the eigenstate $\psi_+$ but is required to make an oscillation, too. Consequently, one can see a rapid transition into the initial state at the end of each period. In Fig.1 (b) and (d), the system is in the $\mathcal{PT}$-broken region and the initial states evolve to $\psi_+$ at large times as expected. \\
{\it{State conversion of the unstable eigenstate}}: In time-independent Hermitian systems, if the initial state is an eigenstate of the system, then it remains in that eigenstate. In non-Hermitian systems, this is not always true \cite{cemmod} as unavoidable noises in the system leads to modulational instability. If the initial state is the eigenstate $\ds{{\Psi_1}}$, then it remains in that eigenstate. However, an initial $\ds{{\Psi_0}}$ is not robust as instability occurs under an infinitesimal perturbation. More specifically, it remains in that eigenstate when $t<t_c$ and then makes a rapid transition into the least decaying eigenstate. The Fig.2 plots the ratio $b/a$ when the system is initially assumed to be in the eigenstate $\ds{{\Psi_0}}$ with small perturbation. The state conversion occurs after some time. As the perturbation gets stronger, the state conversion occurs more rapidly. As can be seen from (a), the EP has nothing to do with the state conversion. 
\begin{figure}[t]
\includegraphics[width=4.25cm]{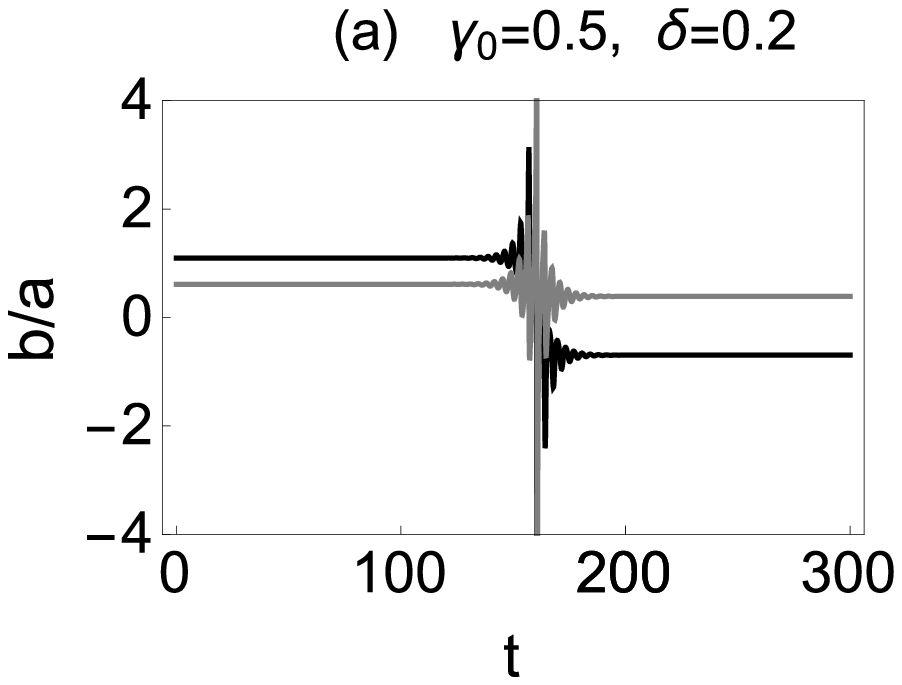}
\includegraphics[width=4.25cm]{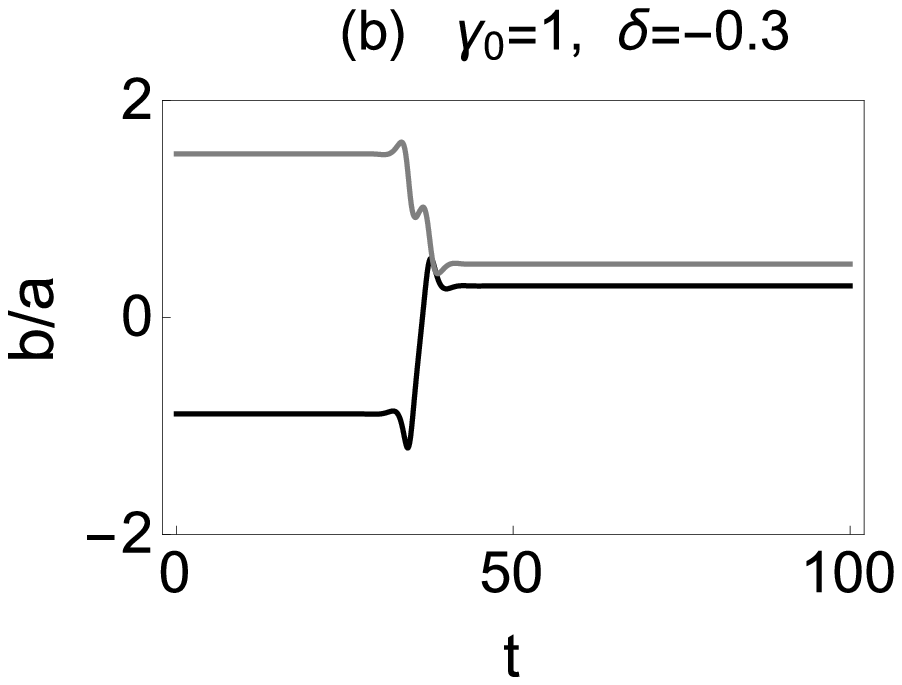}
\caption{ The real (black) and imaginary (gray) parts of the ratio $\ds{b(t)/a(t)}$. The system is initially prepared in the eigenstate $\Psi_0$. Note that $\Psi_0$ is $\psi_-$ when $\delta_0>0$ and $\psi_+$ when $\delta_0<0$ (if $\gamma_0>0$). We add very small amplitude modulation on the initial state and we see that the system keeps its initial state for a while and then makes a rapid transition into the eigenstate $\Psi_1$. }
\end{figure}

\section{Adiabatic evolution}

So far, we have studied the time-independent system. We are now in position to study the time-dependent system. Assume that the parameters $\ds{\gamma(t)}$ and $\ds{\delta(t)}$ change slowly in time. Let us explore adiabatic evolution in this regime. The instantaneous eigenstates for given $\ds{\gamma(t)}$ and $\ds{\delta(t)}$ are defined as $\ds{\psi_\mp(t)=\left(\begin{array}{cc}  1 \\ \delta+i\gamma\pm  \sqrt{1+(\delta+i\gamma)^2}    \end{array}\right)}$ (\ref{oawipjj2}). As we did before, we define $\Psi_1(t)$ and $\Psi_0(t)$ as modulationally stable and unstable instantaneous eigenstate that has higher (lower) imaginary part of instantaneous energy eigenvalues, respectively. Let us write two important conditions in non-Hermitian extension of adiabatic evolution in $\mathcal{PT}$-broken region
\begin{itemize}
  \item  Adiabatic evolution: {\it{The system remains in its instantaneous eigenstate iff the instantaneous eigenstate is modulationally stable. Otherwise, state conversion always occurs.}}\\
  \item Relaxation towards adiabatic evolution: {\it{Any initial state collapses into the modulationally stable instantaneous eigenstate. Therefore adiabatic evolution occurs in the system (just after the state conversion), regardless of the initial state provided that the modulationally stable instantaneous eigenstate is fixed in time.}}\\
\end{itemize}
The former statement is similar to the one in Hermitian system with the exception that not all instantaneous eigenstates but only the modulationally stable instantaneous eigenstate can be adiabatically followed. The latter statement is unique to non-Hermitian system. In Hermitian system, one must start with the instantaneous eigenstate to apply the adiabatic theorem. However, adiabatic theorem can be applied for any initial state in non-Hermitian systems. If the system parameters are varied slowly, then the system follows the modulationally stable instantaneous eigenstate sooner or later.

\subsection{Closed trajectories in the parameter space}

Suppose $\ds{\gamma(t)}$ and $\ds{\delta(t)}$ change  periodically in time with a constant angular frequency (They form a closed loop in the parameter space). There are two characteristic time scales in the system. These are the period $T$ and relaxation time $\ds{t_c}$. We consider the low frequency regime where $\ds{T>>t_c}$ and assume that the initial state is arbitrary. Since the frequency is low, the system has enough time in one cycle for a state conversion into $\Psi_1(t)$, which implies that the initial state becomes unimportant in the dynamics of the system before the system makes one cycle. One can say that adiabatic evolution occurs if there is no more state conversion in the system, which is possible if $\ds{\gamma(t)}$ and $\ds{\delta(t)}$ have either the same or opposite signs at all time. In this case, the system remains in $\Psi_1(t)$ for $t>t_c$. On the other hand, if $\ds{\gamma(t)}$ and $\ds{\delta(t)}$ change their signs relatively, adiabatic evolution is impossible since state conversion between $\psi_+$ and $\psi_-$ occurs periodically in each cycle.\\
The simplest closed trajectory in the parameter space is the circular loop where $\ds{\gamma(t)}$ and $\ds{\delta(t)}$ change periodically in time with period $T=2\pi/\omega$.  
\begin{equation}\label{ypoks2}
\gamma(t)=\gamma_0-\rho \cos{(\omega t-\phi_0)} ,~\delta(t)=\delta_0+\rho \sin{(\omega t-\phi_0)}
\end{equation}
where $|\rho|<|\gamma_0|$ is the radius of the loop, $\ds{(\gamma_0-\rho,\delta_0)}$ is the center of the circular trajectory in the parameter space, $\omega$ is the angular frequency and $\phi_0$ is the initial phase. Note that positive and negative values of the angular frequency denote clockwise (c.w.) and counterclockwise (c.c.w.) circling directions, respectively. Depending on the specific values of $\gamma_0$ and $\rho$, the EP can either be inside or outside of the trajectory. In order to explore topological feature of the dynamics, we will start with this circular trajectory and then consider deformed trajectories. \\
Let us define the complex fidelity for the system with time-dependent state $\psi(t)=\{a(t),b(t)\}^T$ as
\begin{equation}\label{fidel}
F_{\mp}    =\frac{b/a}{\delta(t)+i\gamma(t)\pm  \sqrt{1+(\delta(t)+i\gamma(t))^2}   }
\end{equation}
where the denominator is also equal the ratio $\ds{ b/a  } $ for the instantaneous states. If $F_{\mp}=1$, then the system is in its  instantaneous state $\psi_{\mp}$. In Fig. 3, we plot the fidelities for 3 different trajectories as a function of $\tau=2{\pi}t/\omega$ at $\omega=0.1$. The trajectories are circular and deformed loops in the $1$st quadrant and elliptic loop in the $4$th quadrant in the parameter space as shown in Fig. 3 (d). They are chosen in such a way that $\ds{\gamma(t)}$ and $\ds{\delta(t)}$ have either the same or opposite signs at all points on the trajectories, which lead to adiabatic evolution. We assume that the initial state is $\ds{ \psi (t=0)=\{1,0\}^T }$. We see that the corresponding fidelities for all trajectories rapidly become $1$ ($t_c$ is small) and then remain to be equal to $1$. Consequently, we reach two important conclusions. Firstly, circling on the loop is topological in the sense that any deformation of the loop does not change the result as long as the deformed loop stay in the same quadrant in the parameter space. Secondly, an initial state eventually reach the adiabatically evolved state. Any initial state in such systems follows the adiabatic solution sooner or later, which is a unique feature of non-Hermitian systems. Note that the fidelity would be $1$ for all time if the initial state is the modulationally stable instantaneous state.  \\
A question arises. What happens if $\ds{\gamma(t)}$ and $\ds{\delta(t)}$ change fast, i. e., $\ds{T<<t_c}$. Does the system still follow the modulationally stable instantaneous eigenstate $\Psi_1$? The answer is No as expected. In Fig. 4, we plot the fidelities for a circular trajectory depicted in the $1$st quadrant in Fig. 3 (d) at large $\omega$. In this regime, the system doesn't follow $\Psi_1$ adiabatically. 
\begin{figure}[t]
\includegraphics[width=4.25cm]{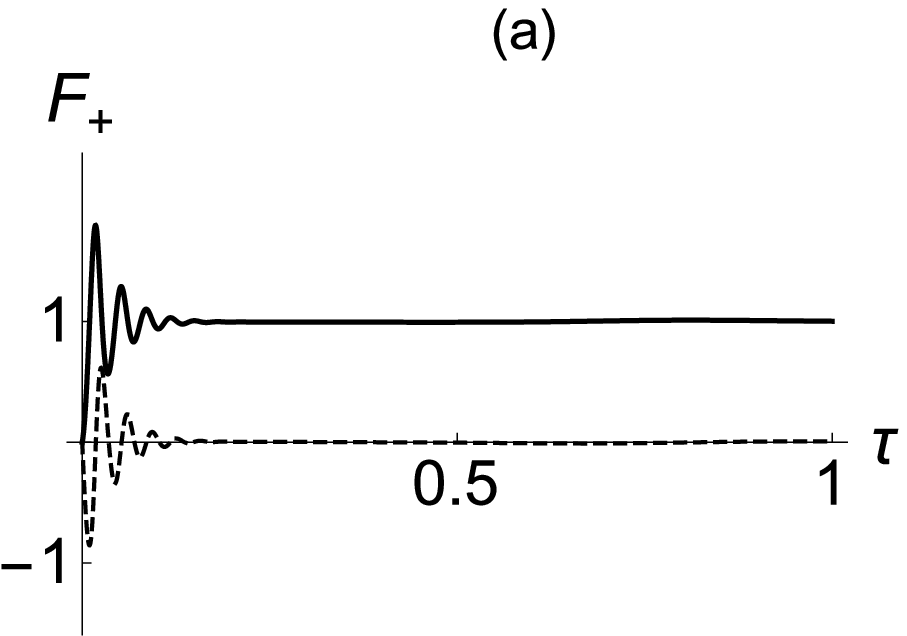}
\includegraphics[width=4.15cm]{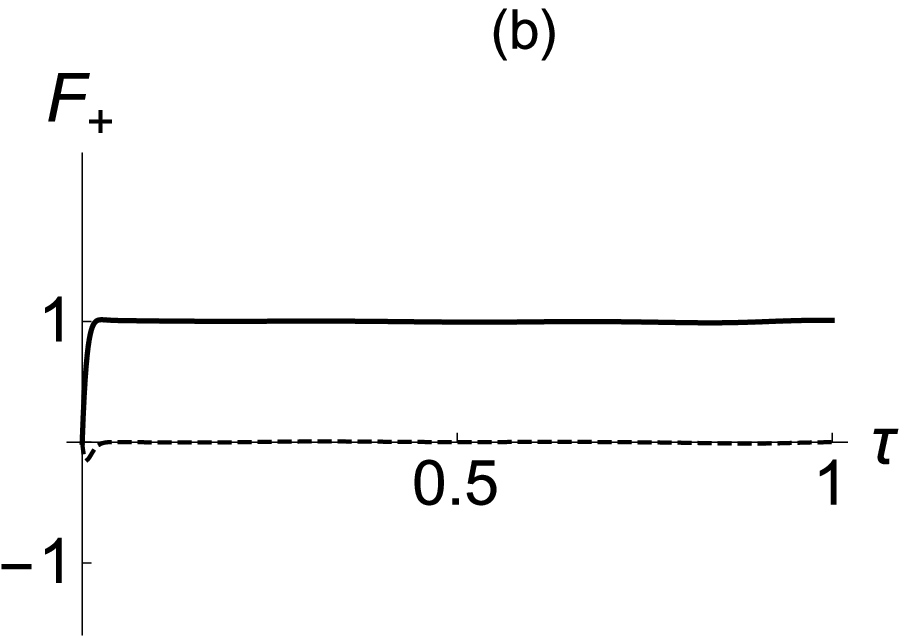}
\includegraphics[width=4.15cm]{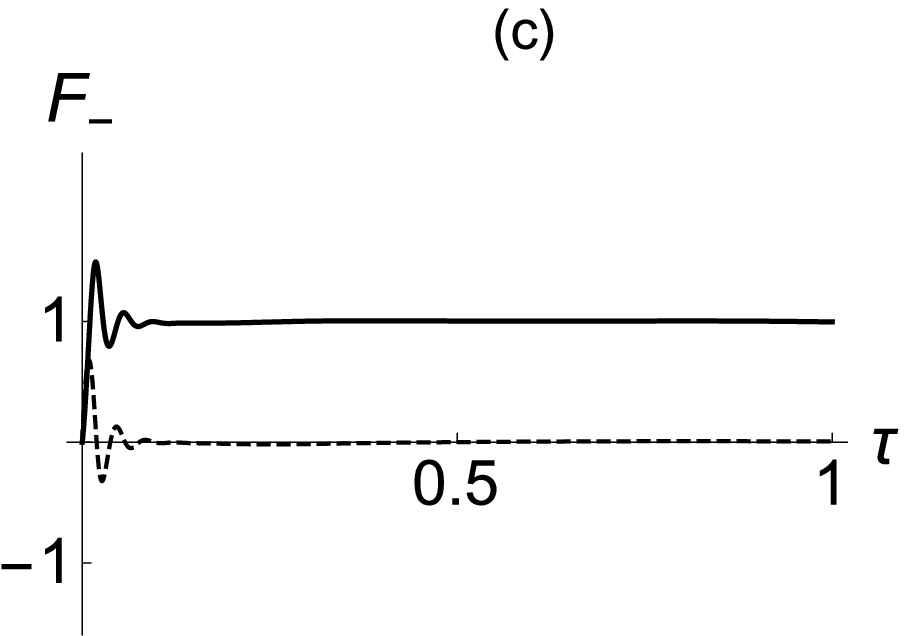}
\includegraphics[width=4.4cm]{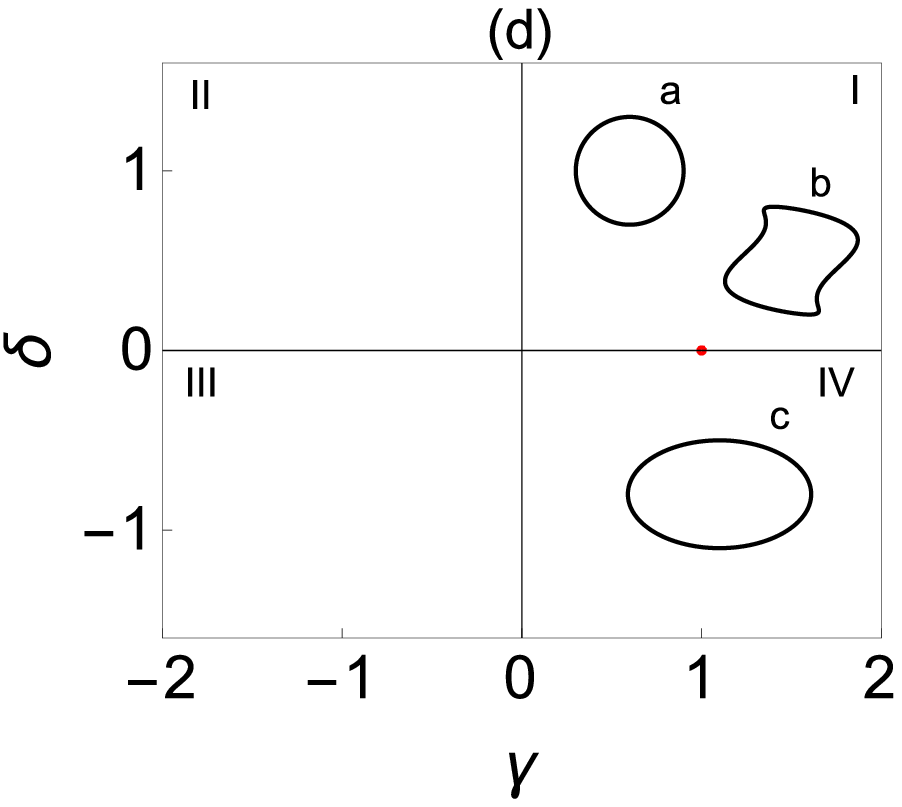}
\caption{ The real (solid) and imaginary (dashed) parts of the fidelities for a circular (a) and deformed (b) trajectories in the $1$st quadrant and an elliptic (c) trajectory in the $4$th quadrant as a function of $\tau=2{\pi}t/\omega$ at $\omega=0.1$. The trajectories are shown in (d). The initial state is given by $\{1,0\}^{T}$. The circling directions (clockwise or counterclockwise) play no role and the three plots (a,b,c) remain the same under $\omega\rightarrow-\omega$.}
\end{figure}
\begin{figure}[t]
\includegraphics[width=4.15cm]{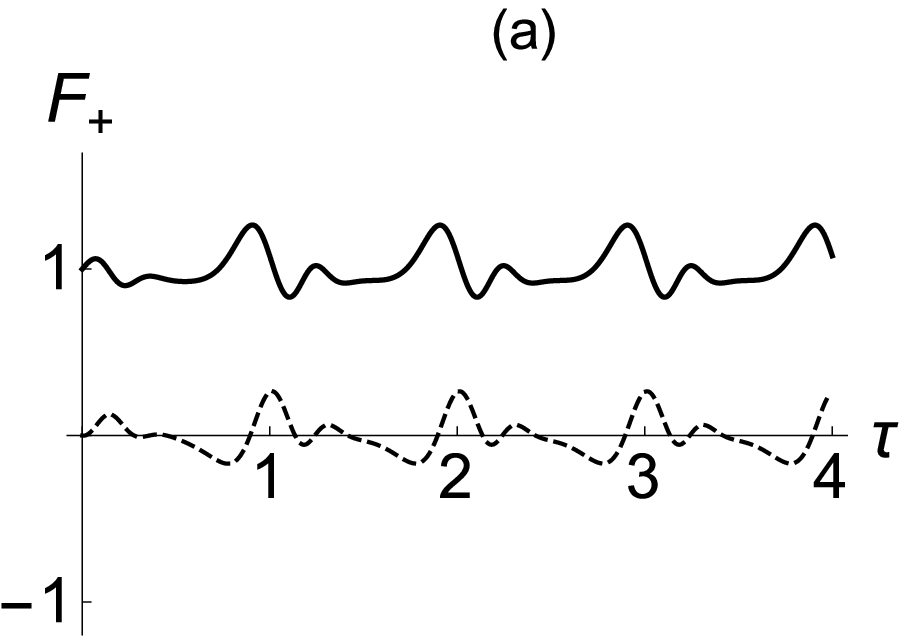}
\includegraphics[width=4.15cm]{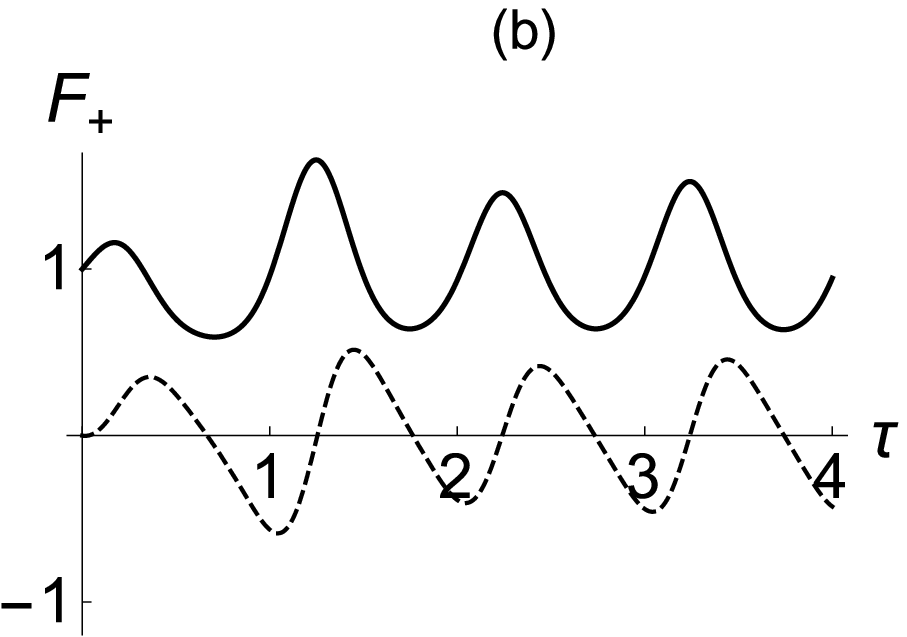}
\caption{ The real (solid) and imaginary (dashed) parts of the fidelities for a circular trajectory in the $1$st quadrant shown in Fig. 3 (d) at $\omega=1$ (a) and $\omega=5$ (b) as a function of $\tau=2{\pi}t/\omega$. The initial state is $\psi_+$. The system does not follow the modulationally stable instantaneous eigenstate when the parameters change fast.}
\end{figure}\\
Suppose next that the slowly changing parameters $\ds{\gamma(t)}$ and $\ds{\delta(t)}$ change their signs relatively during time. In Fig. 5 (d), one can see three such trajectories, where the EP is not encircled. Note that $\psi_+$ ($\psi_-$) is the modulationally stable instantaneous eigenstate in the $1$st and $3$rd ($2$nd and $4$th ) quadrants. In this case, adiabatic evolution is not possible since the modulationally stable instantaneous eigenstate switches between $\ds{ \psi_+ (t) }$ and $\ds{\psi_- (t) }$. The sign of $\omega$ and the initial phase $\phi_0$ play important roles in the dynamics since there is competition between $\ds{{\psi_+}}$ and $\ds{{\psi_-}}$. For example, consider the circular trajectories in Fig 5 (d) with $\phi_0=0$. The system first collapses into $\psi_+$ after relaxation and then state conversion into $\ds{\psi_-}$ occurs for clockwise direction. This is reversed for counterclockwise direction. In Fig. 5 (a-c), we plot the fidelities for each trajectories for the initial state $\{1,0\}^{T}$ and the initial phase $\ds{\phi_0=0}$ at $\omega=0.1$. In  Fig. 5 (a), the trajectory is elliptical and placed mostly in the first quarter where $\psi_+(t)$ is the stable instantaneous eigenstate. Therefore the fidelity $F_+$ becomes $1$ at $t_c\sim 0.4$. After relaxation time, no other state conversion occurs. This is because the system spends less time in the second quarter than the time required for state conversion. In Fig 5 (b) and (c), we plot the fidelities for the two circular trajectories excluding the EP. One of them is near to the origin, which shows that it has higher relaxation time. Note that as the imaginary part of energy eigenvalue increases,  the relaxation time is decreased and state conversion becomes more visible. This can be seen by comparing Fig 5 (b) and (c). In the latter one, relaxation time is smaller and $F_-$ alternates between $1$ and $0$, which shows that the system collapses into $\psi_-$ and $\psi_+$ in an alternating way, regardless of the initial state. Note that this is reversed if we change the circling direction.\\
Finally let us consider closed loops including the exceptional point. In this case, one can certainly say that $\ds{\gamma(t)}$ and $\ds{\delta(t)}$ change their signs relatively during time. In Fig. 6 (d), we see $3$ such loops. The relative change of their signs happens $2$ and $4$ times in each cycle for the loops (a,c) and (b), respectively. As a result, state exchange between the instantaneous states occurs repeatedly. This means that adiabatic evolution is not possible if the EP is encircled. This can be seen from the Fig 6 (a,b,c) since the corresponding fidelities vary in time. The initial state is not important after relaxation time but the initial phase and the circling direction play an important role in the dynamics since they determine the exact position on the loop at which the state conversion occurs. Let us compare the findings in Fig. 5 and Fig 6. We see similar results, which imply that not the inclusion or exclusion of the EP but the exchange of modulationally stable state plays important roles in the dynamics. If the modulationally stable state is fixed in time, then adiabatic evolution is possible even if the initial state is not an instantaneous eigenstate. If it is not fixed in time, state exchange depending on the initial phase and the circling direction occurs. 
\begin{figure}[t]
\includegraphics[width=4.15cm]{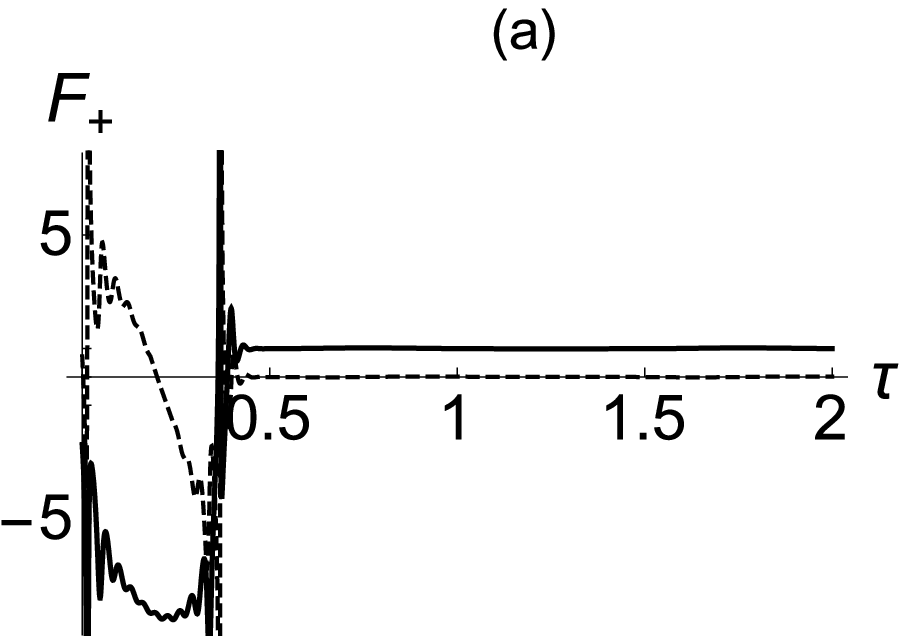}
\includegraphics[width=4.15cm]{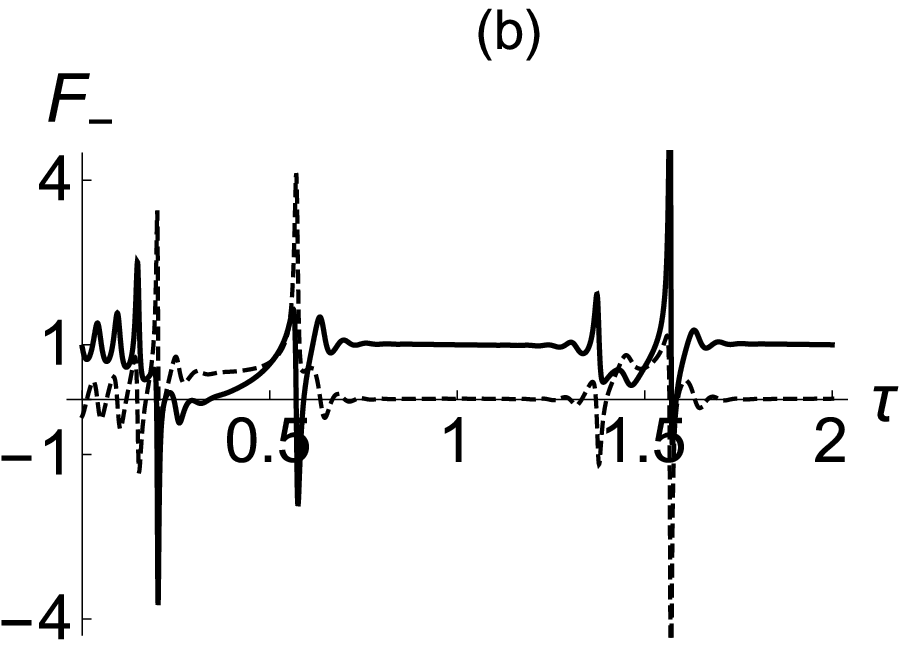}
\includegraphics[width=4.25cm]{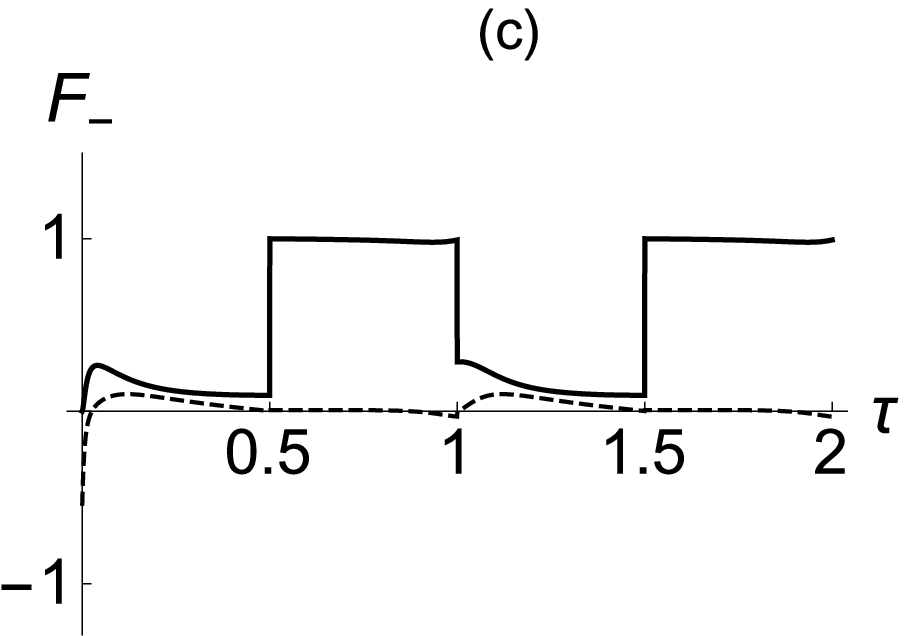}
\includegraphics[width=4cm]{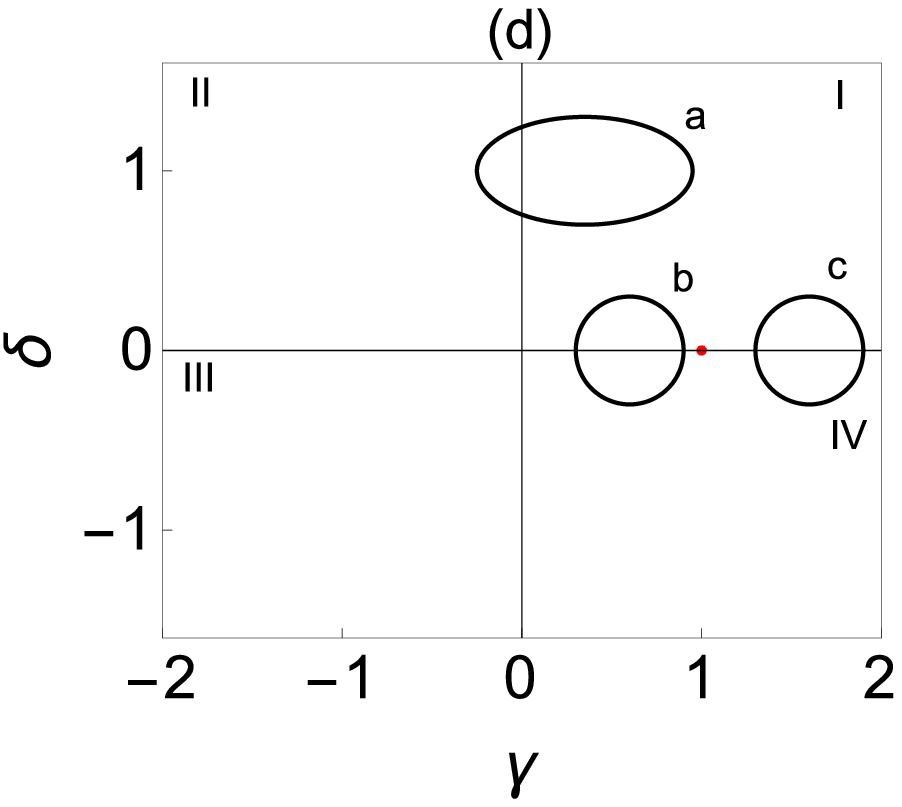}
\caption{ The real (solid) and imaginary (dashed) parts of the fidelities for an elliptical (a) and circular (b,c) trajectories as a function of $\tau=2{\pi}t/\omega$ at $\omega=0.1$. The three trajectories are shown in (d). State conversion is more visible in (c) since the corresponding imaginary part of instantaneous energy eigenvalue is highest among them. We stress that the EP is excluded from all trajectories. The final state becomes $\psi_+$ in (a) for either c.w. or c.c.w. directions since the loop is mostly in the $1$st quadrant. However, this is not the case in (b) and (c) and changing the circling direction changes the final state.}
\end{figure}
\begin{figure}[t]
\includegraphics[width=4.2cm]{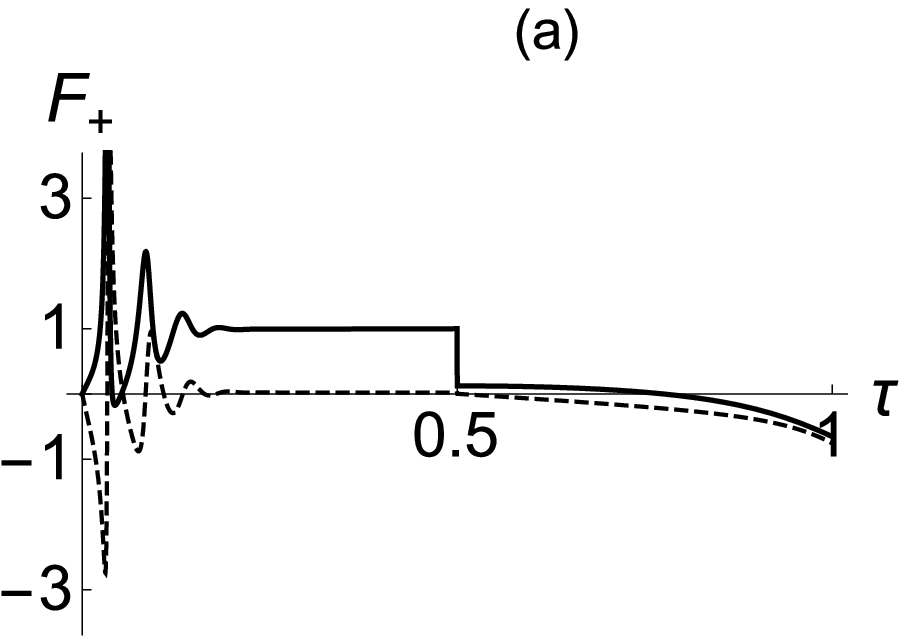}
\includegraphics[width=4.2cm]{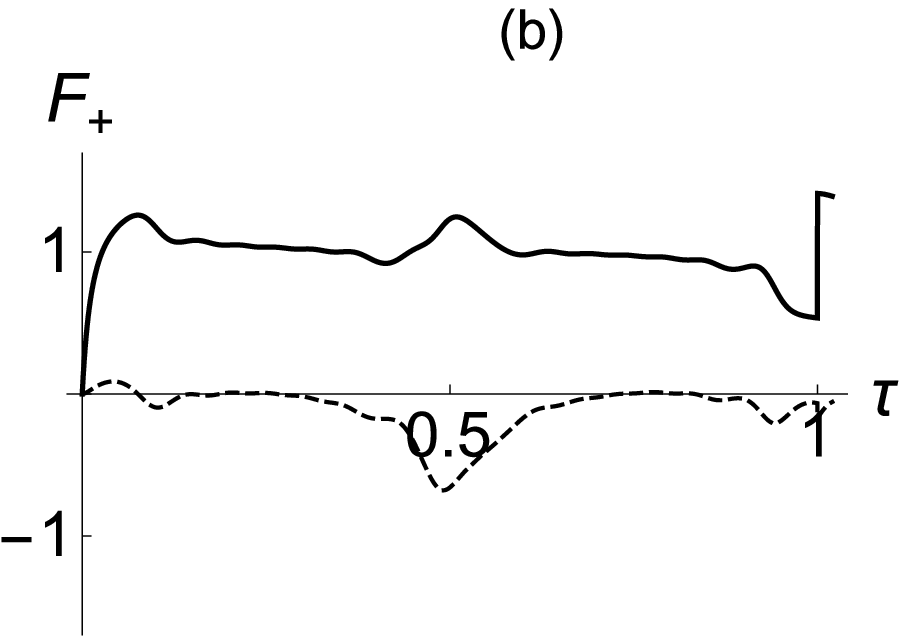}
\includegraphics[width=4.25cm]{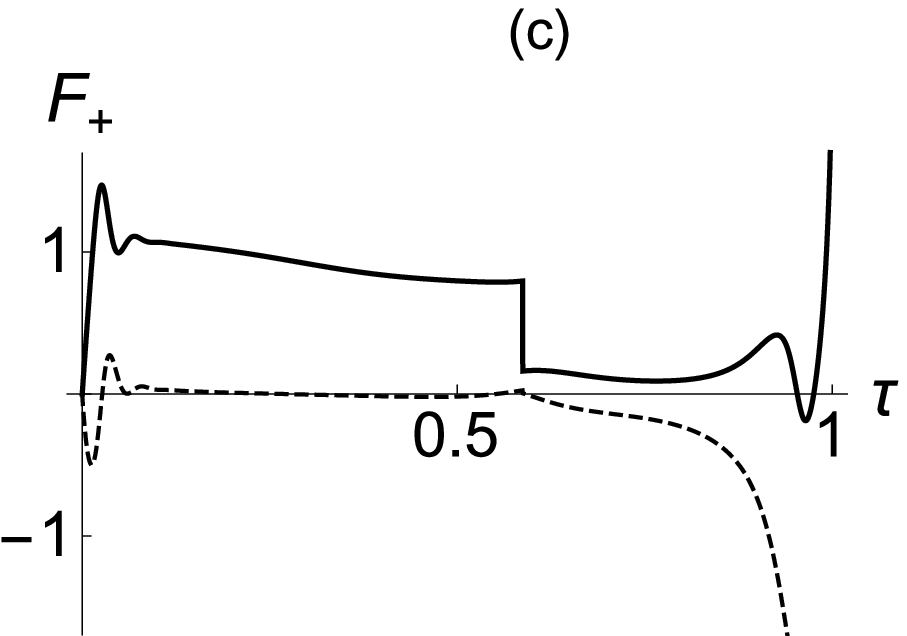}
\includegraphics[width=4cm]{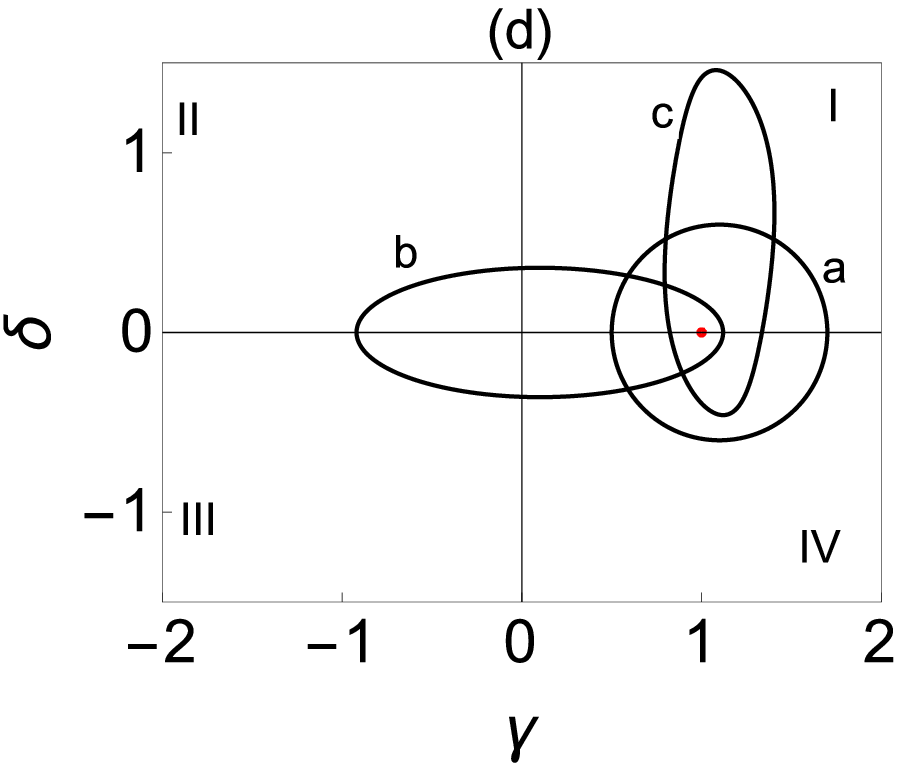}
\caption{The real (solid) and imaginary (dashed) parts of the fidelities for 3 different loops including EP as a function of $\tau=2{\pi}t/\omega$ at $\omega=0.1$. Comparing them with the plots in Fig. (4), we see that adiabatic evolution is not possible if the EP is encircled. This is because of the fact that modulatioanally stable state is not fixed for loops including the EP.}
\end{figure}\\
{\textit{High frequency regime}}: We have explored adiabatic evolution and state conversion in the low frequency regime. Let us now study what happens if the system parameters change fast in a circular loop ($T>>t_c$). In this case, the system comes to its initial position on the circular loop before the state conversion occurs. Furthermore, the system has not enough time to adapt itself to follow the instantaneous eigenstate at any instant. Therefore, it is a good idea to consider the average values of the time-dependent parameters in one period. They are given by $\ds{\overline{ \gamma(t) }=\gamma_0}$ and $\ds{\overline{ \delta(t) }=\delta_0}$. We expect that exclusion or inclusion of the EP plays an important role in the dynamics. We propose to start with the solution (\ref{okjhgs2}) to study the dynamics in the high frequency regime. We claim that a small amplitude oscillation with angular frequency $\omega$ on the solution (\ref{okjhgs2}) occurs. Furthermore, we expect that the amplitude of this oscillation decreases with increasing $\omega$ and decreasing $\rho$. To confirm our qualitative predictions, we perform numerical computations. The Fig. 7 plots the ratio $\ds{b/a}$ for various values of $\gamma_0$ and $\rho$ at $\ds{\omega=2\pi}$. As can be seen, the numerical calculations confirm our predictions. The numerical value of $\ds{\gamma_0}$ plays a central role. If $\ds{\gamma_0=1}$ and $\delta=0$ as in (a,b), the system makes oscillations around the exceptional state. The amplitude of the oscillation increases with $\rho$ as expected. If $\ds{\gamma_0=2}$ and $\delta=0$ as in (c,d), the oscillation occurs around the solution $\psi_+$. We stress that the long time dynamics is independent of the initial states. 
\begin{figure}[t]
\includegraphics[width=4.25cm]{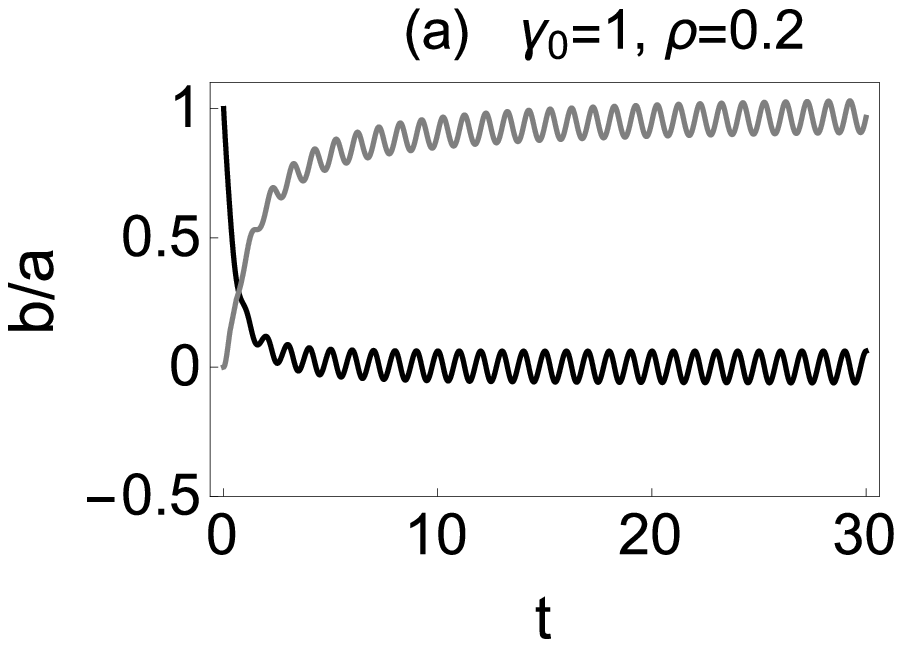}
\includegraphics[width=4.25cm]{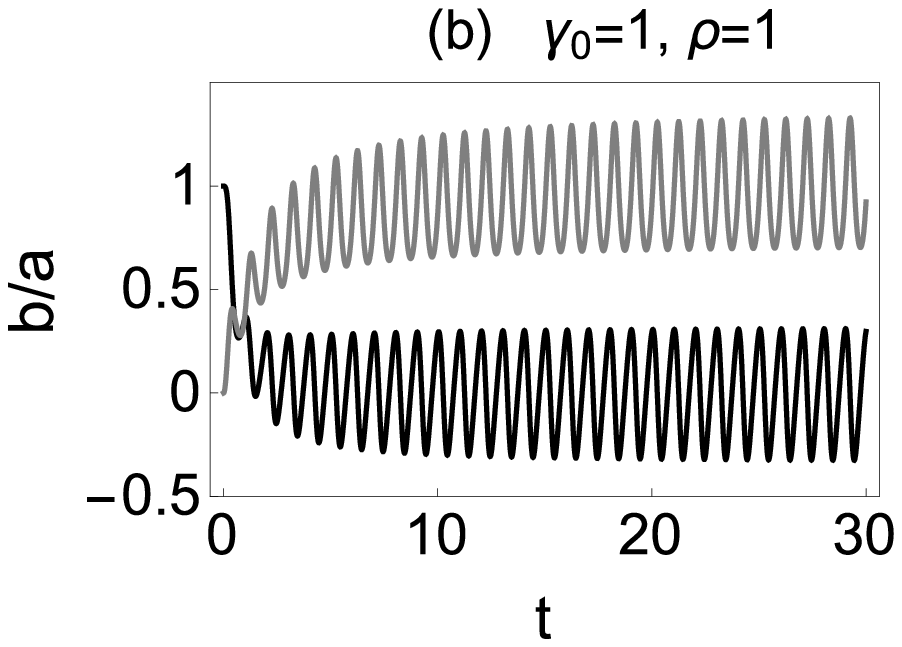}
\includegraphics[width=4.25cm]{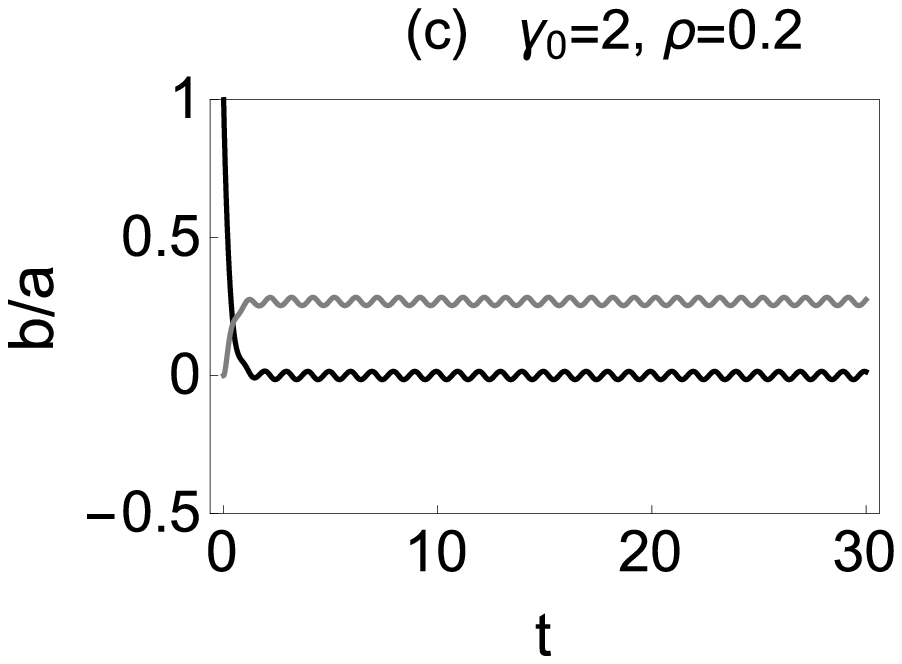}
\includegraphics[width=4.25cm]{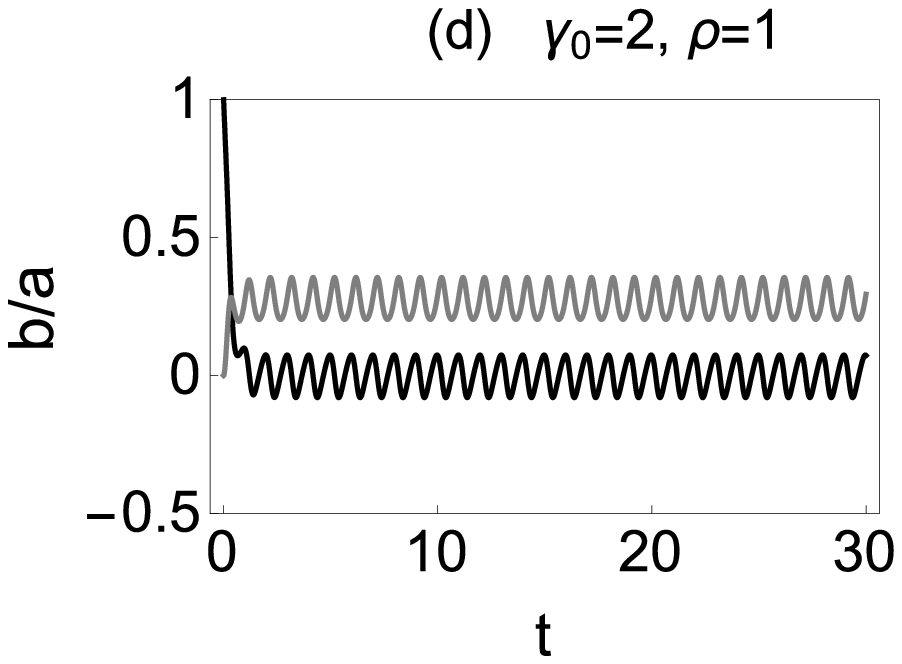}
\caption{ The real (black) and imaginary (gray) parts of the ratio $\ds{b/a}$ at $\ds{\omega=2 \pi}$ for the initial state $\ds{ |\psi(0)>=\{1,1\}^{T}  }$. At large times, the system makes oscillations around the stable eigenstate $\Psi_1$, regardless of the initial states. As can be seen, the amplitude of the oscillation decreases with $\gamma_0$ but increases with $\rho$.}
\end{figure}

\section{Discucssion}

In this paper, we have considered the time-dependent $2\times2$ non-Hermitian Hamiltonian. Consider a general time-independent $N{\times}N$ non-Hermitian Hamiltonian. The system may have $N^{\prime}$th order exceptional points, where $N^{\prime}{\leq} N$. The non-Hermitian Hamiltonian have $N-N^{\prime}$ distinct energy eigenvalues at given parameters. Consider the state $\psi$ that has the highest imaginary part of energy eigenvalues. Then any initial state will evolve to this state. If there are two such states, there exists oscillation between them. If there is another state whose imaginary part of energy eigenvalues is close to that of $\psi$, then it takes more time for the system to evolve to $\psi$ for any arbitrary state.\\
We have discussed that modulationally stable eigenstate is preferred in time, regardless of initial states. Unfortunately, this state grows exponentially in time (the imaginary part of its energy eigenvalue is positive). This may pose some problems for some practical applications.  Let us consider the following loss-dominant Hamiltonian 
\begin{equation}\label{ya0qj2}
\mathcal{H}^{\prime}=\left(\begin{array}{ccccc}i ~(~\gamma_0-|E_{+,I}|~)+\delta_0& -1  \\ -1  & -i(~\gamma_0+|E_{+,I}~)-\delta_0 \end{array}\right)
\end{equation}
which is $\ds{\mathcal{H}^{\prime}=- i|E_{+,I}|~ \mathcal{I}+\mathcal{H}}$, where $\ds{\mathcal{H} }$ is given by (\ref{yudj2}), $|E_{+,I}|$ is the positive imaginary part of $E_+$ and $\mathcal{I}$ is the unitary matrix. The exact solution $\psi(t) $ in (\ref{okjhgs2}) is transformed as $   \exp{(-|E_{+,I}| t)} ~\psi(t)$. Using the large time approximations $\sinh({\nu}   t)\sim \exp({{\nu}   t)}/2i $ and $\cosh({\nu}   t)\sim \exp({{\nu}   t)}/2 $ in (\ref{okjhgs2}), we find that $|\psi(t)>  $ neither grows nor decays in time for large values of time. \\
\textit{  $\mathcal{PT}$ symmetric region: The eigenstates have real valued eigenvalues. Therefore they neither grow nor decay in time. However, any superposition of the eigenstates leads to the so-called power oscillation effect (the total density oscillates in time)\\
$\mathcal{PT}$-broken region: The eigenstates have complex valued eigenvalues. The system collapses into the modulationally stable eigenstate at large times, regardless of initial states. If the highest imaginary part of energy eigenvalues is zero for a loss-dominant Hamiltonian, the total density decays initially and then remains constant, regardless of initial state. }

\section{Conclusion}

In this paper, we have studied time evolution of an arbitrary state in a non-Hermitian system. We have particularly studied dynamic encirclement of exceptional point and analyzed state conversions around exceptional points. We have discussed that initial states are not important in the dynamics of the system in the $\mathcal{PT}$ symmetry broken region. We have shown that the state at large times becomes always the modulationally stable instantaneous eigenstate, regardless of the initial state. We have analyzed adiabatic conditions in our system. On page 3, we give two statements for the non-Hermitian extension of adiabatic evolution. In Hermitian system, one must start with the instantaneous eigenstate to apply the adiabatic theorem. However, adiabatic theorem can be applied for any initial state in non-Hermitian systems. If the system parameters are varied slowly, then the system follows the modulationally stable instantaneous eigenstate sooner or later.\\


\begin{thebibliography}{0}
\bibitem{EP1} T. Kato, Perturbation Theory for Linear Operators (Springer-Verlag, Berlin, 1966).
\bibitem{EP2} Ingrid Rotter, J. Phys. A: Math. Theor. {\bf 42}, 153001 (2009).
\bibitem{EP3} W. D. Heiss, J. Phys. A: Math. Theor. {\bf 45}, 444016 (2012). 
\bibitem{unitran1} H. Ramezani, H.-K. Li, Y. Wang, and X. Zhang, Phys. Rev. Lett. {\bf 113}, 263905 (2014).
\bibitem{unitran2} Z. Lin, H. Ramezani, T. Eichelkraut, T. Kottos, H. Cao, and D.N. Christodoulides, Phys. Rev. Lett. {\bf 106}, 213901 (2011).
\bibitem{unitran3} S. Longhi, J. Phys. A  {\bf 44}, 485302 (2011).
\bibitem{unitran4} Z. J. Wong, Y. L. Xu, J. Kim, K. O'Brien, Y. Wang, L. Feng, and X. Zhang, Nat. Photon. {\bf 10}, 796 (2016).
\bibitem{sense1} W. Chen, S. K. Ozdemir, G. Zhao, J. Wiersig, and L. Yang, Nature {\bf 548}, 192 (2017).
\bibitem{sense2} J. Wiersig, Phys. Rev. Lett. {\bf 112}, 203901 (2014).
\bibitem{listop} T. Goldzak, A. A. Mailybaev, and N. Moiseyev, Phys. Rev. Lett. {\bf 120}, 013901 (2018).
 \bibitem{EPdeney1} C. Dembowski, H.-D. Graf, H. L. Harney, A. Heine, W. D. Heiss, H. Rehfeld, and A. Richter, Phys. Rev. Lett. {\bf 86}, 787 (2001).
\bibitem{EPdeney2} Jorg Doppler, et. al., Nature {\bf 537}, 76 (2016).
\bibitem{EPdeney3} Sang-Bum Lee, et. al., Phys. Rev. Lett. {\bf 103}, 134101 (2009).
\bibitem{EPdeney4} Kun Ding, Guancong Ma, Z.Q. Zhang, and C. T. Chan, Phys. Rev. Lett. {\bf 121}, 085702 (2018).
\bibitem{EPdeney5} H. Xu, D. Mason, Luyao Jiang, J. G. E. Harris, Nature  {\bf 537}, 80 (2016). 
\bibitem{EPmulti1} Jung-Wan Ryu, Soo-Young Lee, Sang Wook Kim, Phys. Rev. A {\bf 85}, 042101 (2012).
\bibitem{EPmulti2} Gilles Demange, Eva-Maria Graefe, J. Phys. A {\bf 45}, 025303 (2012) .
\bibitem{EPmulti3} Jing, S. K. Ozdemir, H. Lu and Franco Nori , Sci. Rep. {\bf 7}, 3386 (2017).
\bibitem{EPmulti4} C Yuce, H Ramezani, arXiv:1812.02218 (2018).
\bibitem{nonh2} C. Yuce, Phys. Lett. A {\bf 379}, 1213 (2015).
\bibitem{nonh3} S Weimann, et. al., Nature Materials {\bf 16}, 433  (2017).
\bibitem{ndiakl38} A Ghatak, T Das, J. Phys. Condens. Matter {\bf 31}, 263001 (2019).
\bibitem{EPatom} Holger Cartarius, Jorg Main, and Gunter Wunner, Phys. Rev. Lett. {\bf 99}, 173003 (2007).
\bibitem{dynexcep2} Absar U. Hassan, Bo Zhen, Marin Soljacic, Mercedeh Khajavikhan, and Demetrios N. Christodoulides, Phys. Rev. Lett. {\bf 118}, 093002 (2017).
\bibitem{encircek0} Dashiell Halpern, Huanan Li, and Tsampikos Kottos, Phys. Rev. A {\bf 97}, 042119 (2018).
\bibitem{encircek1} Hailong Wang, Li-Jun Lang, and Y. D. Chong, Phys. Rev. A {\bf 98}, 012119 (2018).
\bibitem{encirc1} Xu-Lin Zhang, Tianshu Jiang, Hong-Bo Sun, C. T. Chan, arXiv:1806.07649 (2018).
\bibitem{encirc2a} Henri Menke, Marcel Klett, Holger Cartarius, Jorg Main, and Gunter Wunner, Phys. Rev. A {\bf 93}, 013401 (2016).
\bibitem{encirc3a} Ido Gilary, Alexei A. Mailybaev, and Nimrod Moiseyev, Phys. Rev. A {\bf 88}, 010102(R) (2013). 
\bibitem{encirc4a} Qi Zhong, Mercedeh Khajavikhan, Demetrios Christodoulides, Ramy El-Ganainy, Nat. Commun. {\bf 9}, 4808 (2019). 
\bibitem{optoep} H. Xu, D. Mason, L. Jiang, and J. Harris, Nature {\bf 537}, 80 (2016)
\bibitem{sorun1} Absar U. Hassan, Gisela L. Galmiche, Gal Harari, Patrick LiKamWa, Mercedeh Khajavikhan, Mordechai Segev, and Demetrios N. Christodoulides, Phys. Rev. A {\bf 96}, 052129 (2017).
\bibitem{sorun2} X. L. Zhang, S. B. Wang, B. Hou, and C. T. Chan, Phys. Rev. X {\bf 8}, 021066 (2018).
\bibitem{cemmod} C. Yuce, Phys. Rev. A {\bf 99}, 032109 (2019).

\end{thebibliography}
\end{document}